\documentclass[prd,twocolumn,nofootinbib,reprint]{revtex4-1}
\usepackage{amsmath,amssymb,amsthm,bm,braket,ascmac}
\usepackage{graphicx}
\usepackage{color} 

\theoremstyle{definition}

\begin{document}

\title{LHC Future  Prospects of the 750 GeV Resonance}
\author{Ryosuke Sato$^1$ and Kohsaku Tobioka$^{1,2}$}
\affiliation{\vspace{2mm} $^1$Department of Particle Physics and Astrophysics, Weizmann Institute of Science, Rehovot 7610001, Israel \\
$^2$Raymond and Beverly Sackler School of Physics and Astronomy,
Tel-Aviv University, Tel-Aviv 69978, Israel}

\begin{abstract}
\vspace{1mm}
A quantitative discussion on the future prospects of the 750 GeV resonance at the LHC experiment is given using a simple effective field theory analysis.
The relative size of two  effective operators relevant to diphoton decays can be probed by ratios of diboson signals in a robust way. 
We obtain  the future sensitivities of $Z\gamma $, $ZZ$ and $WW$ resonance searches at the high luminosity LHC, rescaling from the current sensitivities at $\sqrt{s}=13$~TeV.  Then, we show that a large fraction of parameter space in the effective field theory will be covered with 300~fb$^{-1}$
and almost the whole parameter space will be tested with 3000~fb$^{-1}$.
This discussion is independent of production processes, other decay modes and total decay width. 
\end{abstract}

\maketitle

\section{Introduction}\label{sec:intro}
Recently, the LHC experiment reported an event excess in diphoton invariant mass distribution
near 750 GeV \cite{ATLAS-CONF-2015-081, CMS-PAS-EXO-15-004, ATLAS-CONF-2016-018, CMS-PAS-EXO-16-018}.
This event excess triggered significant theoretical interest and hundreds of papers have appeared.
One of the plausible candidates to explain this excess is a heavy pion associated with new strong dynamics \cite{Harigaya:2015ezk, Nakai:2015ptz}
(see also Refs.~\cite{
Franceschini:2015kwy,
Low:2015qep,
Bellazzini:2015nxw,
Matsuzaki:2015che,
No:2015bsn,
Bian:2015kjt,
Bai:2015nbs,
Belyaev:2015hgo,
Craig:2015lra, 
Buttazzo:2016kid}).
This model naturally explains the relatively large cross section of the diphoton signal,
and the heavy pion $\phi$ couples with the standard model gauge bosons due to the chiral anomaly as
$\phi B_{\mu\nu} \tilde B^{\mu\nu}$ and $\phi W_{\mu\nu} \tilde W^{\mu\nu}$.
Moreover, it is worth noting that such a model predicts other resonances at the TeV scale, \textit{e.g.}, a color-octet pion as discussed in Ref.~\cite{Bai:2016czm}.
Although such new resonance search is model dependent discussion, 
if we focus on the 750 GeV resonance, a model independent discussion becomes possible.
Actually, there are various models that can be described by a similar effective Lagrangian.
See Ref.~\cite{Franceschini:2016gxv} and references therein.
The effective Lagrangian tells us that we can expect a $Z\gamma $ or $ZZ$ resonance at 750 GeV in addition to the diphoton signal
because of electroweak gauge invariance  \cite{Low:2015qho}.
Also, if some part of diphoton signal is contributed by the $SU(2)_L$ gauge boson, we can expect a $WW$ signal.
Thus it is interesting to discuss the future prospects of diphoton and other modes,
\textit{e.g.}, at the LHC \cite{Fichet:2016pvq, Howe:2016mfq, Chala:2016mdz, Djouadi:2016eyy} and the ILC \cite{Ito:2016kvw}.
In particular, the results of the resonance searches for diboson modes of the LHC 13 TeV run were reported recently \cite{ATLAS-CONF-2016-010, ATLAS-CONF-2016-012, ATLAS-CONF-2016-021, ATLAS-CONF-2015-071, CMS-PAS-EXO-16-019}.
Then it is possible to discuss the future prospects of the diboson modes quantitatively.

In this paper, we discuss the future prospects of the 750 GeV resonance by using an effective field theory analysis.
We take the effective Lagrangian for the 750 GeV resonance and calculate the ratios of the signal strengths for diboson modes\footnote{
A recent trigger level analysis of dijet resonance search gives an upper bound of $\sim3~{\rm pb}$ on 750 GeV resonance
if we take 40 \% acceptance \cite{ATLAS:2016xiv}.
Although we do not discuss dijet search, this could be important in some models.
}.
These ratios are free from QCD corrections and it could tell us some information about the UV theory behind the 750 GeV resonance.
For example, in the case of the heavy pion models, the charge of the ``new quarks'' can be extracted from the ratio of signal strengths.
This information will be useful for the discrimination of the various models.
Here we discuss the high luminosity LHC with $\sqrt{s}=13$~TeV for simplicity. However  results we obtain in this paper can be applied to those at $\sqrt{s}=14$~TeV because parton luminosities ($gg, \bar{q}q$) at 750~GeV are changed by only up to 25\% \cite{Martin:2009iq}.

\section{Analysis}\label{sec:analysis}
For concreteness, we assume that the 750 GeV diphoton resonance is explained by a CP odd scalar boson $\phi$.
The effective interaction of $\phi$ with electroweak gauge bosons is
\begin{align}
{\cal L}_{\rm eff}
= &
\frac{k_Y}{m_\phi}\frac{\alpha_Y}{4\pi} \phi B_{\mu\nu} \tilde B^{\mu\nu}
+ \frac{k_L}{m_\phi}\frac{\alpha_2}{4\pi} \phi W_{\mu\nu} \tilde W^{\mu\nu},
\end{align}
where $k_L$ and $k_Y$ are determined by the UV theory.
As we mentioned in the introduction, this is an effective theory of various models.
If we consider a CP-even scalar,
the effective interactions are changed to $\phi B_{\mu\nu} B^{\mu\nu}$ and $\phi W_{\mu\nu} W^{\mu\nu}$
and the following discussion is same in the limit of $m_\phi \gg m_{Z,W}$ assuming no mixing with the Higgs boson.
By using the above effective interaction, the partial widths of $\phi$ decaying into a pair of electroweak gauge bosons are given as
\begin{align}
&\Gamma(\phi\to \gamma\gamma) = \frac{\alpha^2(k_Y+k_L)^2 m_\phi}{64\pi^3},\\
&\Gamma(\phi\to Z\gamma )     = \frac{\alpha^2(k_Y t_W -k_L t_W^{-1})^2 m_\phi }{32\pi^3} \left(1-\frac{m_Z^2}{m_\phi^2}\right)^3,\\
&\Gamma(\phi\to ZZ)           = \frac{\alpha^2(k_Y t_W^2+k_L t_W^{-2})^2 m_\phi }{64\pi^3} \left(1-\frac{4m_Z^2}{m_\phi^2}\right)^{3/2},\\
&\Gamma(\phi\to W^+W^-)       = \frac{\alpha^2k_L^2 s_W^{-4} m_\phi}{32\pi^3}  \left(1-\frac{4m_W^2}{m_\phi^2}\right)^{3/2},
\end{align}
where $s_W = \sin\theta_W$ and $t_W = \tan\theta_W$ with the Weinberg angle $\theta_W$.

\begin{figure}[t!]
\centering
\includegraphics[width=0.9\linewidth]{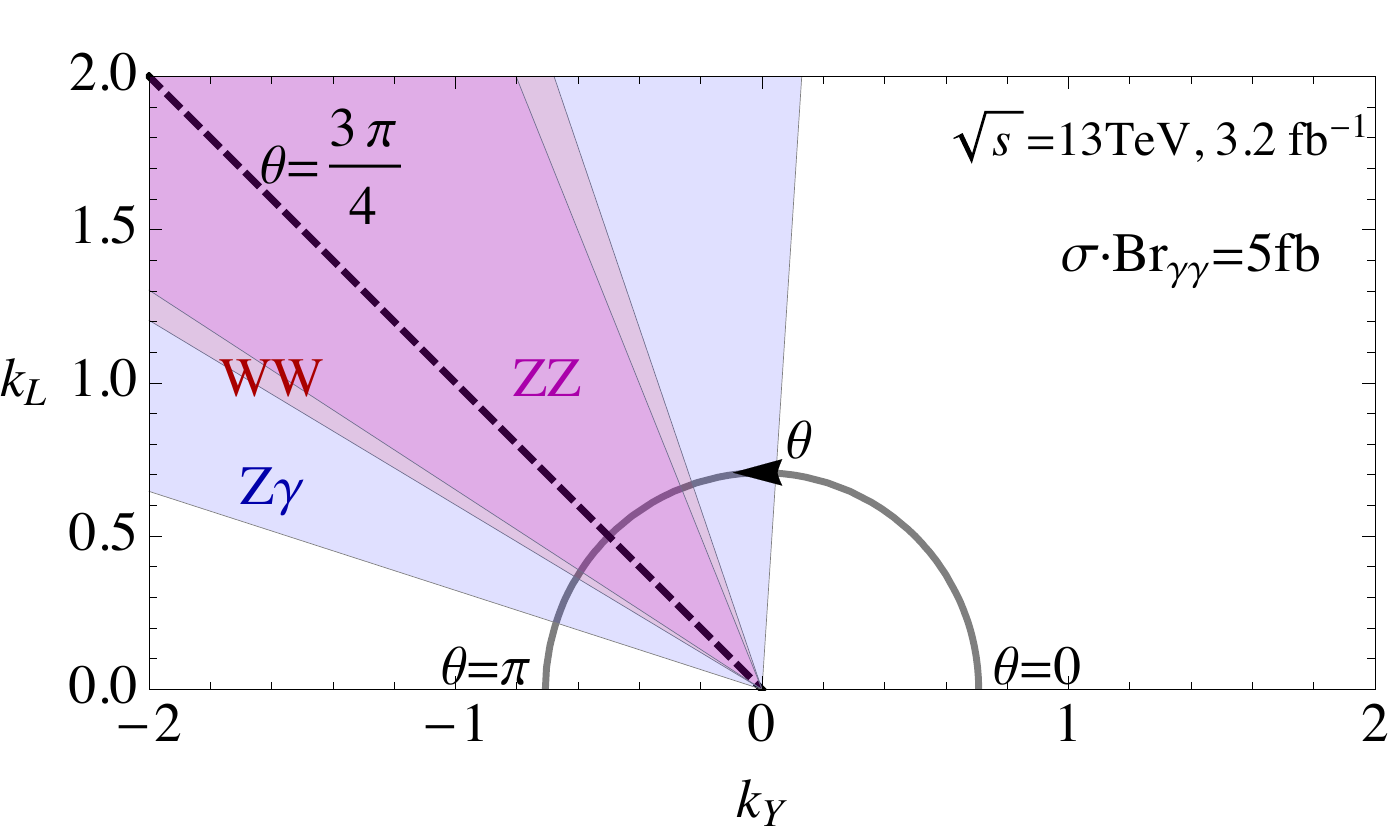}
\caption{ 
Current exclusion limits  at $\sqrt{s}=13$ TeV in $k_Y$-$k_L$ plane. For a fixed diphoton rate, $\sigma\cdot{\rm Br}_{\gamma\gamma}=5~{\rm fb}$, shaded regions are excluded by  $Z\gamma $ (blue) \cite{ATLAS-CONF-2016-010}, $ZZ$ (magenta) \cite{ATLAS-CONF-2016-012}, and $WW$ (red) \cite{ATLAS-CONF-2016-021} resonance searches. 
Bounds at $\sqrt{s}=8$ TeV are not considered because  translation to 13~TeV depends on the production process. 
\label{fig:kLkY}
}
\end{figure}

Although there are four possible observables of $\sigma\cdot{\rm Br}$ in the electroweak gauge boson final states against two parameters,
a robust measurement is possible only for a relative size of the two parameters $k_Y/k_L$
(see Refs.~\cite{Craig:2015lra, Low:2015qho} for related works). 
Therefore, we parametrize the coefficients as \cite{Craig:2015lra}
\begin{align}
k_Y = k\cos\theta, \quad   k_L = k \sin\theta .
\end{align}
By taking ratios of the signal strengths, 
we obtain functions depending on only $\theta$ and $\theta_W$, 
\begin{align}
\frac{\sigma\cdot{\rm Br}_{Z\gamma }}{\sigma\cdot{\rm Br}_{\gamma\gamma}}
	&= \frac{2(t_W \cos\theta - t_W^{-1}\sin\theta)^2}{(\cos\theta+\sin\theta)^2}  \left(1-\frac{m_Z^2}{m_\phi^2}\right)^3,
	\label{eq:gammaZ}\\
\frac{\sigma\cdot{\rm Br}_{Z Z}}{\sigma\cdot{\rm Br}_{\gamma\gamma}} 
	&= \frac{(t_W^2 \cos\theta + t_W^{-2}\sin\theta)^2}{(\cos\theta+\sin\theta)^2} \left(1-\frac{4m_Z^2}{m_\phi^2}\right)^{3/2},
	\label{eq:ZZ}\\
\frac{\sigma\cdot{\rm Br}_{WW}}{\sigma\cdot{\rm Br}_{\gamma\gamma}} 
	&= \frac{2s_W^{-4}\sin^2\theta}{(\cos\theta+\sin\theta)^2} \left(1-\frac{4m_W^2}{m_\phi^2}\right)^{3/2}.
	\label{eq:WW}
\end{align}
We, therefore,  study diboson ($ZZ, Z\gamma , WW$) final states  with respect to $\theta$ and ${\sigma\cdot{\rm Br}_{\gamma\gamma}}$. 
The discussion  is independent of other decay modes, total decay width, and production processes
(therefore independent of QCD uncertainties associated with the productions from quarks and gluons). 
In order to see the above statement more explicitly, it is shown in Fig.~\ref{fig:kLkY} that  the LHC bounds on diboson resonances at 750 GeV are given by the angle $\theta$ and are insensitive to $k$ if we take a fixed value of $\sigma\cdot{\rm Br}_{\gamma\gamma}$.

\begin{table}
\centering
\begin{tabular}{|c|c|c|c|}
\hline
    ${\cal L}$    & $3.2~{\rm fb}^{-1}$(expected)  & $300~{\rm fb}^{-1}$& $3000~{\rm fb}^{-1}$ \\\hline\hline
${\sigma\cdot{\rm Br}}(Z\gamma )$    & 42~fb  \cite{ATLAS-CONF-2016-010}         &  4.3~fb          &  1.4~fb  \\
${\sigma\cdot{\rm Br}}(ZZ)$   & 160~fb \cite{ATLAS-CONF-2016-012}            &  17~fb          &  5.2~fb      \\
${\sigma\cdot{\rm Br}}(WW)$ &  270~fb \cite{ATLAS-CONF-2016-021}          &  28~fb          &  8.8~fb   \\\hline
\end{tabular}
\caption{Future sensitivities (95\% CL) of cross section in $Z\gamma $, $ZZ$, and $WW$ resonances searches with luminosities of 300 and 3000 fb$^{-1}$ at $\sqrt{s}=13~\rm TeV$, rescaled from the  ATLAS expected upper bounds with ${\cal L}=3.2~{\rm fb}^{-1}$ \cite{ATLAS-CONF-2016-010, ATLAS-CONF-2016-012, ATLAS-CONF-2016-021}. 
}\label{tab:prospect}
\end{table}

\begin{figure*}[t!]
\centering
\hspace{-100pt}
\includegraphics[width=0.35\linewidth]{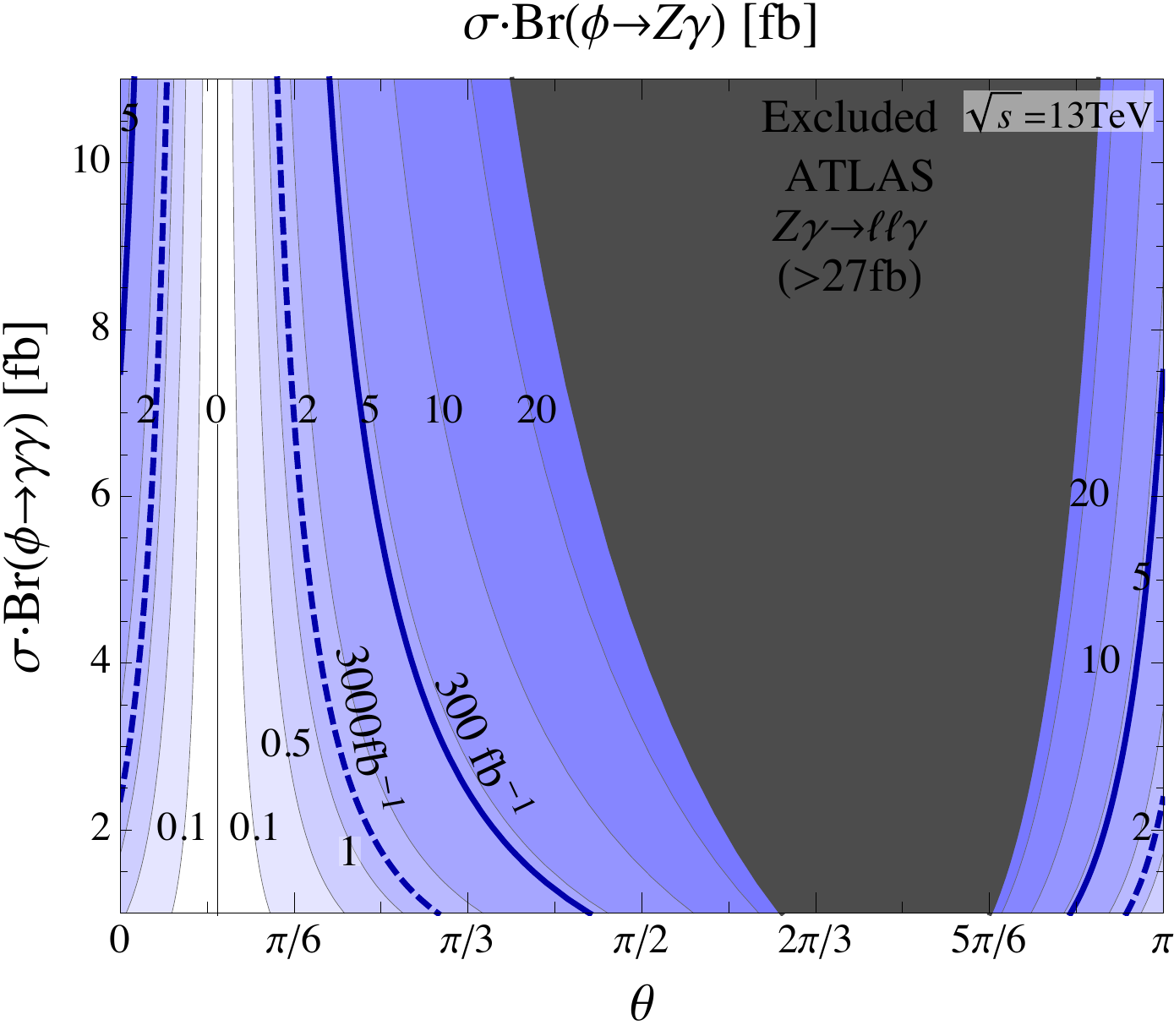}
\includegraphics[width=0.35\linewidth]{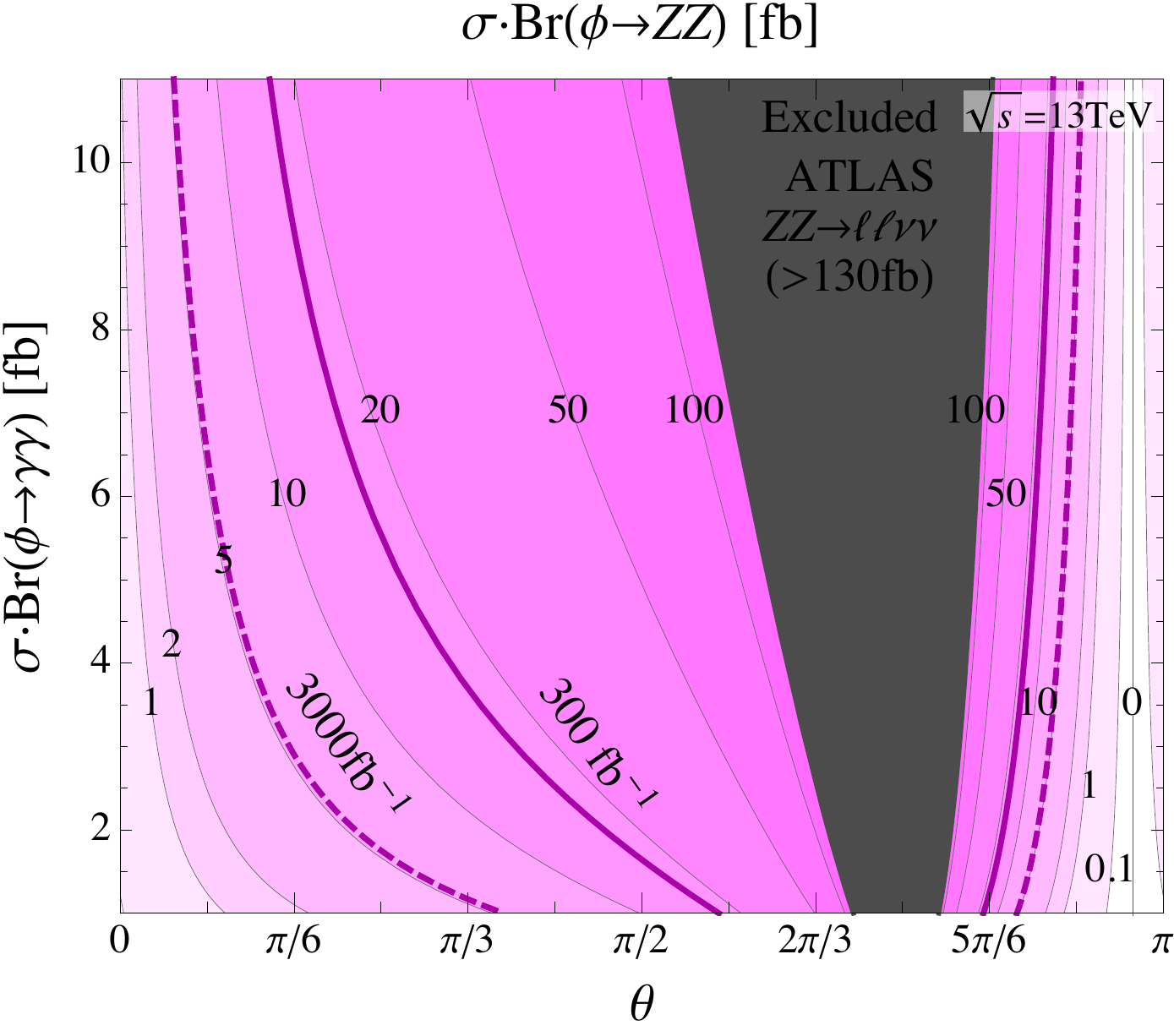}
\includegraphics[width=0.35\linewidth]{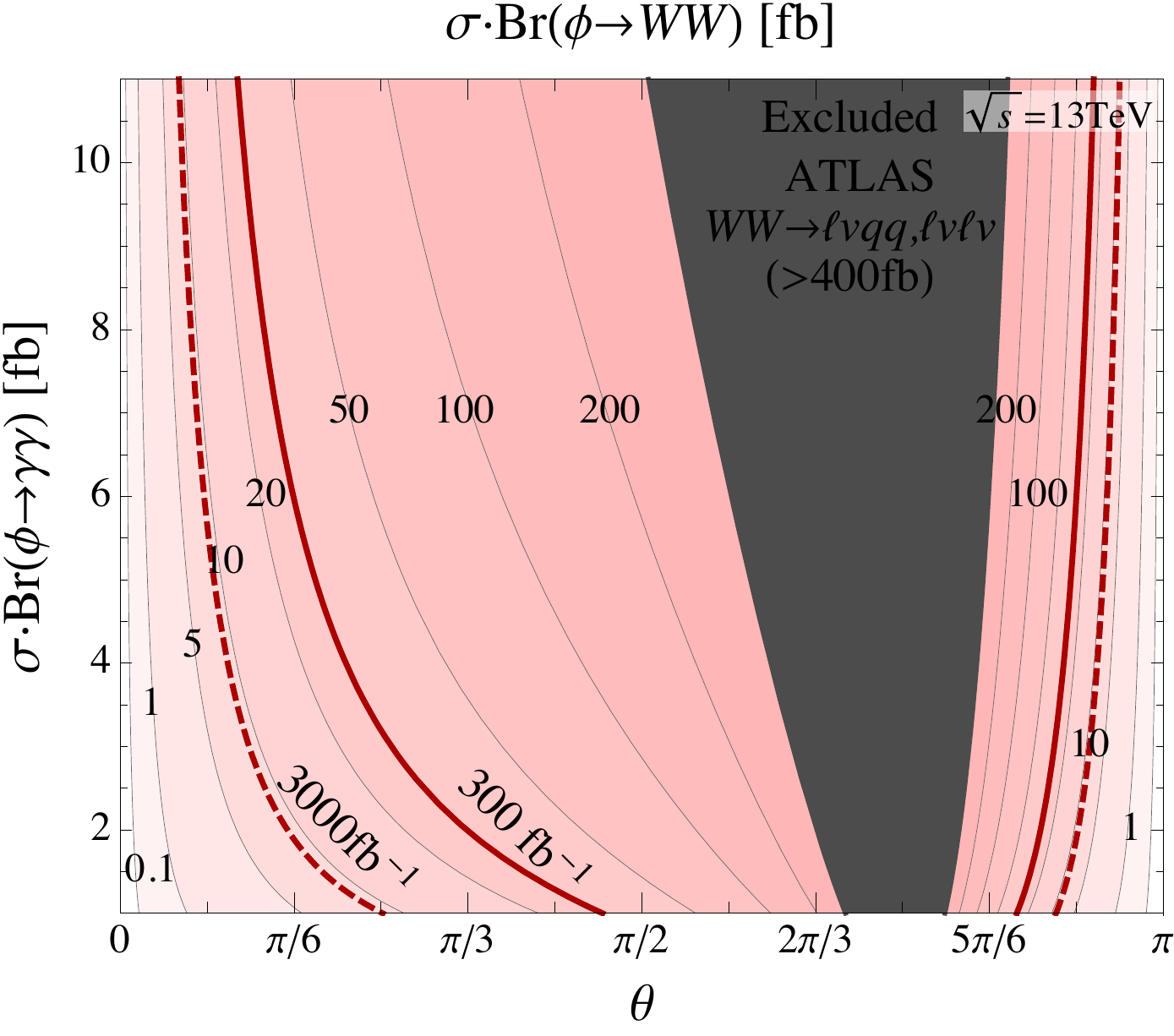}
\hspace{-100pt}
\caption{ Current bound, expected cross section, and future prospects for 750~GeV resonance in $Z\gamma $ ({\it left}), $ZZ$ ({\it center}), and $WW$ ({\it right}) final states. Gray lines show expected cross section at $\sqrt{s}=13$ TeV, and thick  (dashed) lines correspond to expected sensitivity at 95\%~CL with 300(3000)~fb$^{-1}$. Gray  shaded regions are excluded by ATLAS $Z\gamma $ \cite{ATLAS-CONF-2016-010}, $ZZ$ \cite{ATLAS-CONF-2016-012}, and $WW$ \cite{ATLAS-CONF-2016-021} resonance searches with 3.2~fb$^{-1}$. 
\label{fig:prospect}
}
\includegraphics[width=0.48\linewidth]{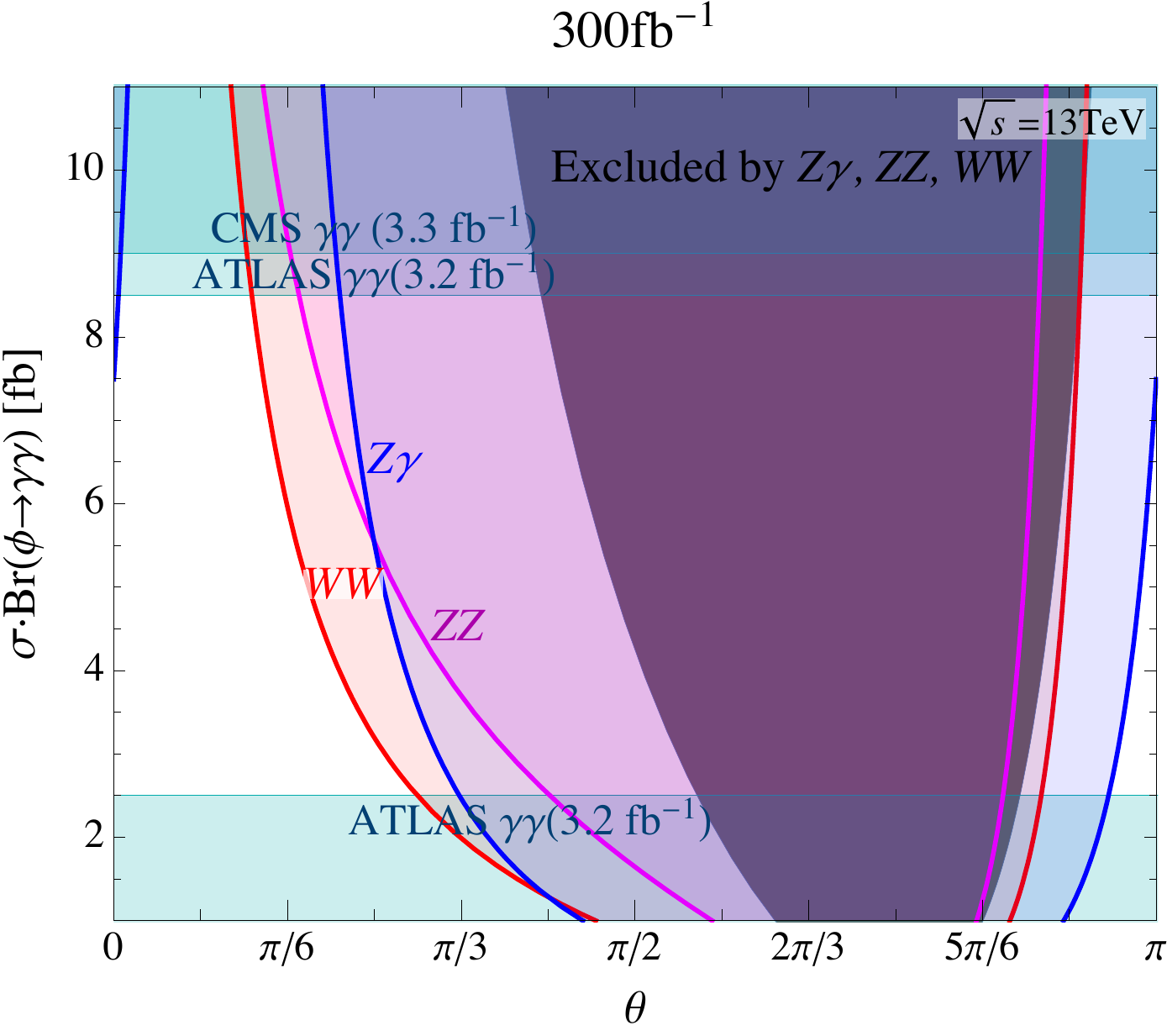}
\hspace{10pt}
\includegraphics[width=0.48\linewidth]{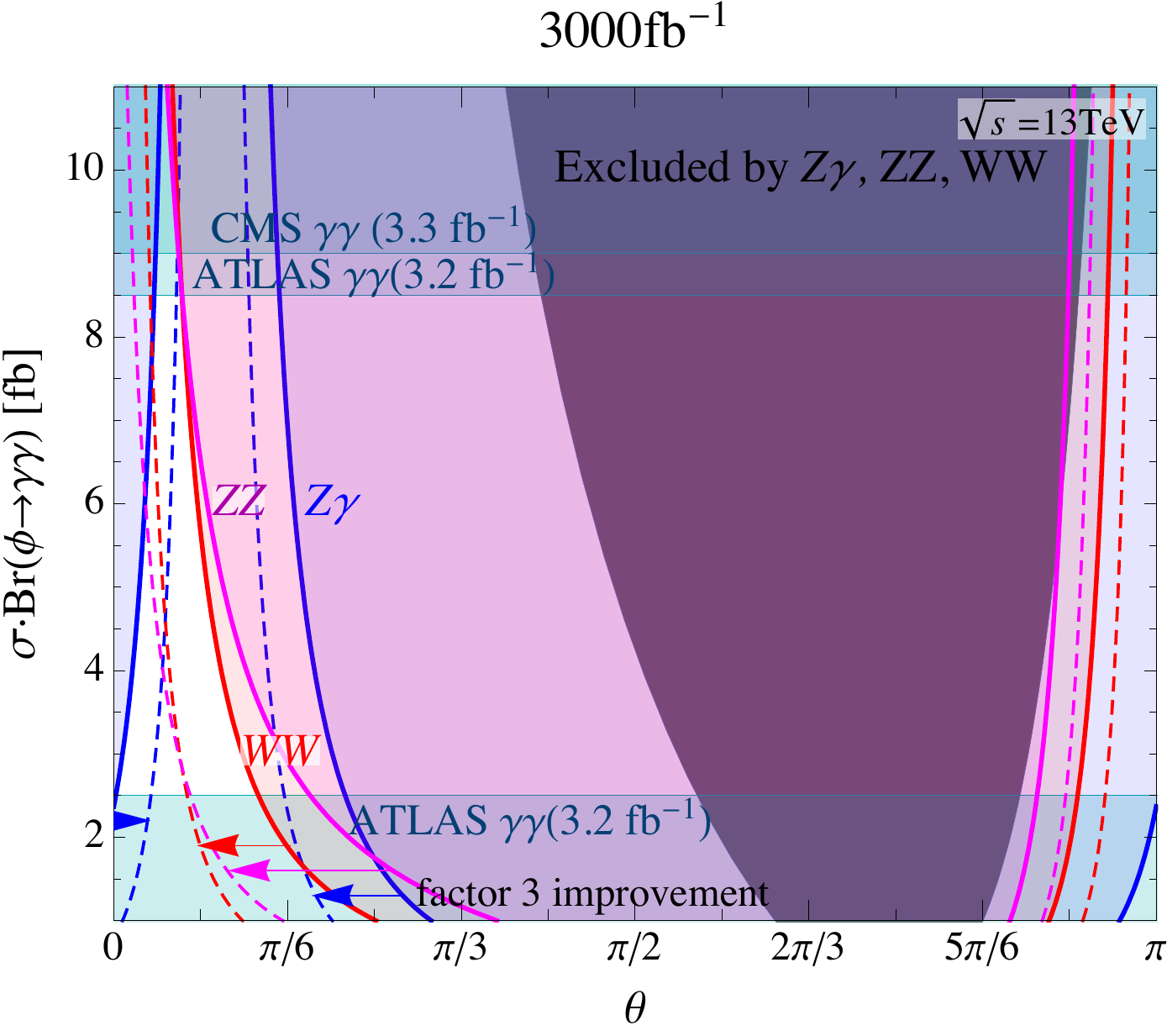}
\caption{ 
Future prospects for 750~GeV resonance in $Z\gamma $ , $ZZ$, and $WW$  final states 
with 300~fb$^{-1}$({\it left}) and 3000~fb$^{-1}$({\it right}).
The shaded regions with magenta, red, and blue lines can be probed in $Z\gamma $ , $ZZ$, and $WW$ channels at 95\%~CL, respectively. 
The upper and lower shaded regions (cyan) are out of 2$\sigma$ allowed region of CMS and ATLAS diphoton searches assuming narrow width \cite{Franceschini:2016gxv}, and the central gray shaded region is excluded by ATLAS diboson resonance searches as in Fig.~\ref{fig:prospect}. The dashed lines in the {\it right} panel correspond to future sensitivities with 3000~fb$^{-1}$  if a factor of 3 improvement is achieved in each channel. 
\label{fig:prospect_combined}
}
\end{figure*}

For the LHC future prospects, based on the current bounds of $3.2~{\rm fb}^{-1}$, 
we obtain future sensitivities in Table~\ref{tab:prospect}, using Gaussian probability distribution. 
We assume a narrow width of the resonance here.
In a case of large width the bounds and sensitivities become weaker by up to 50\% because backgrounds in the wider mass range will be relevant  
(\textit{c.f.}, a difference between narrow width approximation (NWA) and large width approximation (LWA) in Ref.~\cite{ATLAS-CONF-2016-021}). 
For the estimation of the future sensitivities of
the $ZZ$ and $WW$ resonances at $\sqrt{s}=13~\rm TeV$,
we used the $\ell\ell\nu\nu$ \cite{ATLAS-CONF-2016-012} and $\ell\nu qq+ \ell\nu\ell\nu$ \cite{ATLAS-CONF-2016-021} final states, respectively.
For $ZZ$ mode, a search in $\ell\ell q q$ final state \cite{ATLAS-CONF-2015-071} is not included and a search in $4\ell$ final state is not reported yet. For $Z\gamma$ mode,  we adopted ATLAS result \cite{ATLAS-CONF-2016-010} and CMS reported a similar result \cite{CMS-PAS-EXO-16-019}.  
In Fig.~\ref{fig:prospect}, we show current bound, expected cross section, and future prospects for each  diboson resonance search in ($\sigma\cdot{\rm Br}_{\gamma\gamma}$)-$\theta$ plane.  
At angle of $\theta \simeq 3\pi/4$, the cross sections are enhanced because of cancellation in the diphoton channel as in Eqs.~(\ref{eq:gammaZ}, \ref{eq:ZZ}, \ref{eq:WW}). This is also seen in Fig.~\ref{fig:kLkY}. 
On the other hand, there are angles where each diboson channel is cancelled. 
Eqs.~(\ref{eq:gammaZ}, \ref{eq:ZZ}, \ref{eq:WW}) tell that the signals vanish at 
\begin{eqnarray}
\theta=\left\{
\begin{array}{l}  \arctan t_W^2 \simeq 0.09 \pi \\
 -\arctan t_W^4+\pi \simeq 0.97 \pi \\
  0  
\end{array}
\right .
\begin{array}{l}  \text{for }Z\gamma \\
  \text{for }ZZ\\
   \text{for }WW
\end{array}.\ 
\end{eqnarray}
The angles $\theta$ and $\theta + \pi$ in the present effective Lagrangian are physically equivalent,
and we can see the cancellation angles are rather close to each other due to the small Weinberg angle.
Thus we need large luminosity to cover the region near $\theta \simeq 0, \pi$.

The combined plots are presented in Fig.~\ref{fig:prospect_combined}. We also include the 2$\sigma$ allowed range of diphoton rate assuming narrow width \cite{Franceschini:2016gxv}, 
\begin{align}
\quad\quad&2.5\lesssim\sigma\cdot{\rm Br}_{\gamma\gamma} \lesssim 8.5\ {\rm fb} & (\rm ATLAS, 3.2~fb^{-1}),  \nonumber\\
\quad\quad&0.6\lesssim\sigma\cdot{\rm Br}_{\gamma\gamma} \lesssim 9\ {\rm fb}& (\rm CMS, 3.3~fb^{-1}). 
\end{align}
While most of the parameter space will be covered, a region around the $Z\gamma $ cancellation angle of $\theta\simeq 0.09\pi $ cannot be reached with ${\cal L}=300~{\rm fb}^{-1}$ as shown in Fig.~\ref{fig:prospect_combined} left. Part of the region can be covered by the $WW$ channel with the higher luminosity of $3000~{\rm fb}^{-1}$ as shown in Fig.~\ref{fig:prospect_combined} right. However, a small region near $\theta\simeq 0.06\pi $ still remains uncovered. This window can be almost closed if a factor of 3 improvement in sensitivities can be achieved, especially in the $ZZ$ or $WW$ channels. Such improvement could occur by (i) multivariate analysis, (ii) including searches in $4\ell$ and $\ell\ell qq$ final states, (iii) combining ATLAS and CMS data, and (iv) improvement of $Z$- and $W$-taggers. 

Finally, we apply the results to specific models. 
If we take the simplest heavy pion model with $SU(2)_L$ singlet ``new quarks'' (see \textit{e.g.}, Ref.~\cite{Nakai:2015ptz}), 
$\phi$ does not have $\phi W_{\mu\nu}\tilde W^{\mu\nu}$ and then $\theta=0$.
In this case, Fig.~\ref{fig:prospect_combined} shows $Z\gamma $ resonance searches will be important for the future.
Other interesting examples are GUT motivated models.
In a GUT motivated heavy pion model \cite{Harigaya:2015ezk},
$\phi$ is a SM singlet in an adjoint multiplet of $SU(5)_{GUT}$.
The GUT relation predicts $\theta = \arctan 9/5 \simeq 0.34 \pi$,
and Fig.~\ref{fig:prospect_combined} shows $WW$ and $Z\gamma$ resonance searches will be important with 300~fb$^{-1}$ and  $ZZ$ resonance search will be relevant if ${\sigma\cdot{\rm Br}_{\gamma\gamma}}\gtrsim 4$~fb.
If $\phi$ is a singlet of $SU(5)_{GUT}$, the GUT relation predicts $\theta = \arctan 3/5 \simeq 0.17 \pi$ \cite{Craig:2015lra},
and Fig.~\ref{fig:prospect_combined} shows $WW$ resonance search will be important with 300~fb$^{-1}$ and also $ZZ$ resonance search will be relevant with 3000~fb$^{-1}$.
The ratio of the signal strength is important information for the discrimination of the various models.

%

\section{Conclusion}\label{sec:conclusion}
In this paper, we discussed the LHC future prospects of the 750 GeV diphoton resonance.
As pointed out in Refs.~\cite{Craig:2015lra, Low:2015qho},
the ratio of the signal strengths between different final states can be discussed
by using the coefficient of the effective interactions.
Our analysis is based on $\sqrt{s}=13$~TeV, but the results can be applied to those at  $\sqrt{s}=14$~TeV.  We found, at the high luminosity LHC, a large fraction of the parameter space in the effective theory will be covered with 300 fb$^{-1}$
and almost the whole parameter space will be tested with 3000 fb$^{-1}$.
In particular, in the case with only $\phi B_{\mu\nu}\tilde B^{\mu\nu}$ ($\theta = 0$) as in the simplest heavy pion model,
$Z\gamma$ resonance searches will have good sensitivity.
Also, in the GUT motivated case ($\theta = \arctan 9/5 \simeq 0.34 \pi$), 
$WW$ and $Z\gamma$ resonance searches will have good sensitivity  and also $ZZ$ resonance search will be relevant with 300~fb$^{-1}$.


\section*{Acknowledgements}
We thank Elina Fuchs for useful discussions in the early stage of this work.
We also thank Yevgeny Kats and Lorenzo Ubaldi for careful reading and helpful comments on the manuscript.

\bibliography{ref}
\bibliographystyle{utphys}
\end{document}